\documentclass[useAMS]{mn2e}
\usepackage{epsfig}
\title[Period gap LMXBs]{Are the very faint X-ray transients period
gap systems?}

\author[Maccarone et al.]{Thomas J. Maccarone\\ School of Physics and
Astronomy, University of Southampton, Hampshire SO17 1BJ,United
Kingdom\\ \newauthor Alessandro Patruno\\ Sterrenkundig Instituut
Anton Pannekoek, Universiteit van Amsterdam, Postbus 94249, 1090 GE,
Amsterdam, The Netherlands\\}

\begin{document}
\def\ltsim{\mathrel{\rlap{\lower 3pt\hbox{$\sim$}}
        \raise 2.0pt\hbox{$<$}}}
\def\gtsim{\mathrel{\rlap{\lower 3pt\hbox{$\sim$}}
        \raise 2.0pt\hbox{$>$}}}

\date{}

\pagerange{\pageref{firstpage}--\pageref{lastpage}} \pubyear{}

\maketitle

\label{firstpage}

\begin{abstract}
We discuss a scenario for the very faint X-ray transients
 as X-ray binaries fed by winds from detached M-dwarf
donors in binary stars within the ``period gap'' -- the range of
periods where donor stars have become fully convective, and shrunken
so that they no longer fill their Roche lobes, but have not yet
re-attached due to the systems shrinking through gravitational
radiation.  This wind-fed detached binary scenario can reproduce the
two key properties of the very faint X-ray transients -- their
faintness, which defines them, and their relatively low duty cycle
outbursts which require that they have low mean mass transfer rates.
We discuss feasible observational tests of the scenario.

\end{abstract}


\section{Introduction}

The first observations of transient accretion discs -- of the dwarf
nova SS Cyg in outburst -- were reported more than a century ago
(discovered by Louisa Wells, but reported by Pickering \& Fleming
1896).  The mechanism for the transient behaviour is now believed to
be the ionization instability model (Smak 1984).  In this model, the
transient behaviour is proposed to occur due to the fact that the
viscosity of neutral gas is lower than that of ionized gas.  Gas is
heated during the intervals between transient outbursts, and goes
through limit cycles in which it is sometimes hot enough to be
ionized, leading to a rapid viscous timescale, and sometimes cool
enough to be neutral, leading to lower accretion rates.

X-ray binaries also show transient behaviour.  The standard ionization
instability model cannot explain the details of X-ray transient
behaviour (see e.g. Lasota 2001 for a review).  On the other hand, the
effects of irradiation of the outer part of the accretion disc by the
inner, brighter part (e.g. Dubus et al. 1999), and incorporating
irradiation effects seem to allow the basic picture of the ionization
instability to be applied to X-ray binaries with the data matching the
model qualitatively.  A few key specific predictions of the model are
found in the data -- nearly all strong black hole candidates in low
mass X-ray binaries (LMXBs) are observed to be transients, and whether
a neutron star will be a transient correlates well with the
predictions based on observed mean accretion rate and observed orbital
period.

In recent years a new class of X-ray transients has been discovered.
These systems have low peak X-ray luminoisities and short outburst
durations (Sakano et al. 2005; Muno et al. 2005; Wijnands et
al. 2006), earning them the name ``very faint X-ray transients''
(VFXTs).  The VFXTs are problematic for binary stellar evolution.  The
implied mean mass accretion rates from these systems are below about
$10^{-13} M_\odot {\rm yr}^{-1}$, a rate which cannot be reached
through normal Roche lobe overflow processes from a star unless the
accretion rate changes abruptly (King \& Wijnands 2006).  Note that by
``mean mass accretion rate'', we are referring to the mean rate at
which matter is transferred from the donor star to the outer accretion
disk, and that this is best estimated empirically by averaging the
source luminosity over several outburst cycles, and then applying a
conversion between mass accretion rate and X-ray luminosity.

In fact, for cataclysmic variables, there exists a well known
situation in which the accretion rate {\it does} change abruptly --
the period gap.  Cataclysmic variables with initial orbital periods
less than about 10 hours are expected to evolve to shorter orbital
periods through magnetic braking. When the systems reach an orbital
period of about 3 hours, the donor stars are expected to become fully
convective, provided that it still has some significant hydrogen
content (Pylyser \& Savonije 1989; Podsiadlowski et al. 2002), at
which point the star quickly shrinks in size, becomes detached and the
mass transfer rates drop to a sufficiently low level that cataclysmic
variables are not expected to be observed (Warner 1976; Rappaport,
Verbunt \& Joss 1983; Spruit \& Ritter 1983; Patterson 1984).
Eventually, at periods of about 2 hours, gravitational radiation
brings the binary back into contact, and accretion begins again, this
time with the evolution of the binary driven primarily by
gravitational radiation rather than magnetic braking.  In this paper,
we show that the properties of the very faint transients are well
explained if these systems are low mass X-ray binaries in the period
gap, with a low level of accretion from the weak wind of the M-dwarf
donor star.\footnote{We note that a different model which also invoked
wind accretion was proposed for some of the fainter Galactic Centre
region X-ray binaries, by Pfahl et al. (2002).  That model suggested
that the winds were from stars of at least 3 $M_\odot$, and is
unlikely to explain transients which peak at relatively low
luminosities -- but it could explain moderately bright persistent
sources.} We discuss possible observational tests of the scenario.

\section{VFXTs as period gap LMXBs}

From the analogy with cataclysmic variables, it is clear that X-ray
binaries should have a period gap, and that the accretion rates of
systems within the period gap should be much lower than those for
systems at longer or shorter periods.  In fact, Spruit \& Ritter
(1983) noted that there should be a period gap for low mass X-ray
binaries. The more relevant question, then, is whether one can expect
a high enough accretion rate for period gap X-ray binaries that these
systems would be seen in appropriate numbers and whether the expected
properties of the systems would match well to the observed properties
of the systems.

\subsection{Expected outburst peak luminosities}

We can address the latter point first.  The fundamental properties of
very faint X-ray transients is that they have outbursts with low peak
luminosities.  The peak outburst luminosities of X-ray binaries have
been shown to be well correlated with the X-ray binaries' orbital
periods (e.g. Shahbaz et al. 1998; Portegies Zwart et al. 2004; Wu et
al. 2010).  Since the period gap occurs for periods of $\approx2-3$
hours, these will be at the short period end of the distribution of
orbital periods -- these systems should have faint outbursts.

A few different relations can be used to estimate the peak luminosity,
$L_{p}$, expected from a transient with an orbital period of 2-3
hours.  First, we can take a theoretical constraint, from King \&
Ritter (1998).  They suggest that $L_p=2.3 \times10^{36} R_{11}$
erg/sec, where $R_{11}$ is the outer disk radius in units of $10^{11}$
cm.  The binary separation will be about $7.7\times10^{10}$ cm, for an
orbital period of 2.5 hours, if the accretor mass is 1.4$M_\odot$ and
the donor mass is 0.25 $M_\odot$.  The circularization radius of the
accretion disk, under the assumption of Roche lobe overflow, will be
about 0.2 times the binary separation (see below), while the tidal
radius of the accretion disk will be about 0.9 times the radius of the
donor star (FKR02).  These values are the same to within a few
percent, and as we discuss below, even if the wind speed is large, so
that the Roche lobe overflow approximation is a poor one, we can still
expect that the matter in the disk will spread out until it reaches
the tidal radius. The peak $L_X$ can thus be expected to be about
$4\times10^{35}$ erg/sec for a typical period gap system, consistent
with the observations for many VFXTs.

Alternatively, we can take the empirical relations between
$L_p/L_{EDD}$ and orbital period.  Sources in the range from $2-3$
hours in orbital period are found to be at a few percent of the
Eddington luminosity, yielding an expected peak luminosity of a few
times $10^{36}$ ergs/sec (Wu et al. 2010).  However, it should be
noted that the biases in the sample of Wu et al. (2010) are such that
they should push up the luminosities of the short period systems --
the systems were necessarily detected as transients in order to have
been included in the sample, and current all-sky instruments will not
detect very faint X-ray transients at Galactic Center distances.

The outburst durations are more difficult to estimate on theoretical
grounds.  The observed outburst durations of the VFXTs are often short
-- days to weeks (Wijnands 2008).  This is in line with the
statistical study of Portegies Zwart et al. (2004), which found that
the decay times of outbursts should be a few days at orbital periods
of a few hours.  When one looks specifically at the few systems with
orbital periods between 1.4 and 4.5 hours which have well understood
transient outbursts, some systems (e.g. 1A 1744-361) have frequent
short outbursts consistent with our expectations (e.g. Bhattacharyya
et al. 2006), while others have longer outbursts (e.g. EXO~0748-676 --
Degenaar et al. 2009).  We note, on the other hand, that given that in
many cases, these systems should have accretion disks which extend to
or beyond their tidal radii, they may be susceptible to
``super-outbursts'' like those seen in the SU Uma class of cataclysmic
variables (Osaki 1989) -- given observational selection effects that
work strongly against detecting faint, short outbursts, the X-ray
binary equivalent of superoutbursts may represent the dominant mode of
outbursts seen from short period systems by all-sky instruments.  In
SU UMa stars, the superoutbursts are both brighter and longer than
``typical'' dwarf nova outbursts.

\subsection{Expected mass transfer rates}
Next, we can estimate the range of accretion rates possible for these
systems, check whether the systems are likely to be transients, and
then make estimates of the likely duty cycles.  The mass trasnfer
rates we calculate should be $\sim10^{-14}-10^{-15} M_\odot$/yr in
order for the model to work effectively.  If the mass transfer rates
are much larger than $10^{-14} M_\odot$/yr, then the systems will have
duty cycles larger than those observed, and frequent outbursts would
be expected for a large fraction of the sources.  If, on the other
hand, the mass transfer rates are much less than $10^{-15}
M_\odot$/yr, then a very large underlying population of sources would
be needed just to produce the observed transient events.

There are multple possible mechanisms for accretion from M-dwarfs by
neutron stars.  The most likely is that the M-dwarfs have weak, but
non-zero stellar winds which can be captured relatively effectively by
their neutron star companions.  Such winds probably exist as well for
Roche-lobe overflowing stars, but, because of their low rates, they
can be ignored relative to the much larger mass loss rates passed from
the donor star through Roche lobe overflow.  These winds may be
important, on the other hand, for the case of at least some of the low
accretion rate polar systems, cataclysmic variables which show
similarly hard-to-explain low mean accretion rates, one of which is
found in the period gap (e.g. Schwope et al. 2002).

The mass loss rates from M-dwarfs are highly uncertain.  Debes et
al. (2006) find typical values of $\dot{M}$ of about
$10^{-14}-10^{-16} M_\odot$/year by looking at a class of white dwarfs
which have detectable metal lines in their atmospheres, and assuming
that the metals come from accretion of the white dwarf wind.  Models
of cataclysmic variable evolution implictly assume mass loss rates for
M-dwarfs similar to the solar mass loss rate of $2\times10^{-14}
M_\odot$/yr (Sills et al. 2000; Schreiber \& Gaensicke 2003).

Next we can consider the fraction of the mass we can expect to be
accreted by the neutron star companion of an M-dwarf which is almost,
but not qute in Roche lobe contact.  Assuming a wind speed of the
escape velocity from the donor star, the accretion rate due to wind
mass transfer can be found to be:

\begin{equation}
\frac{\dot{M}}{-\dot{M}_W} = \frac{1}{4} \left(\frac{M_{ACC}}{M_{DON}}\right)^2 \left(\frac{R_{DON}}{a}\right)^2
\end{equation}
 where $\dot{M}$ is the accretion rate onto the compact star,
 $-\dot{M}_W$ is the mass loss rate from the donor star, $M_{ACC}$ is
 the mass of the accretor, $M_{DON}$ is the mass of the donor star,
 and $R_{DON}$ is the radius of the donor star (Frank, King \& Raine
 2002).  For systems which are nearly Roche lobe overflowing, this
 number will approach unity.  There are some indications that the
 Bondi-Hoyle rate may overestimate the real wind mass transfer rates
 by a factor of about 10, especially in systems where the inferred
 Bondi rate is a very large fraction of the total mass loss from the
 donor star.  It seems quite reasonable for a substantial fraction of
 M-dwarf/neutron star binaries to have mass transfer rates in the
 $10^{-14}-10^{-16} M_\odot$/year range where the very faint X-ray
 transients are most likely to be found.  Furthermore, there is some
 evidence that the wind mass loss rates of rapidly rotating M-dwarfs
 (for example from looking at the rotation periods of M-dwarfs in
 different Galactic kinematic populations), such as those expected to
 exist in short period binaries may be larger than the typical values
 for more slowly rotating M-dwarfs (e.g. Irwin et al. 2011).

\subsection{Should disks form?}

To test whether accretion disks should form in these wind-fed systems,
one can calculate the circularization radius of an accretion disk in
the case of an M-dwarf in Roche lobe overflow, with an orbital period
of 2.5 hours, and an M-dwarf with a mass of 0.25 $M_\odot$ and a
radius of 0.25 $R_\odot$, and we can compare these circularization
radii.  We note that we have intentionally chosen the radius we use to
be slightly smaller than measured radii of M-dwarfs (Fernandez et
al. 2009) in order to make conservative assumptions about disk
formation.  We can compute the circularization radius, $R_{circ}$ in
terms of the binary separation, $a$ and the binary mass ratio $q$ in
the Roche lobe overflow case by taking equation (4.20) of Frank, King
\& Raine (2002):

\begin{equation}
R_{circ}/a = (1+q) [0.5-0.227 \rm{log}~~q]^4,
\end{equation}

finding that $R_{circ}/a=0.24$ for this case.  We can then take
equation (4.42) of FKR02 and find:

\begin{equation}
\frac{R_{circ}}{a} = \frac{M_N^3 (M_N + M_E)}{16 \lambda^4 M_E^4} \left(\frac{R_E}{a}\right)^4
\end{equation}

where $M_N$ is the mass of the neutron star accretor, $M_E$ is the
mass of the star emitting a wind, and $\lambda$ is the ratio of the
wind speed to the escape velocity from the wind emitter.  Evaluating
with the values given above, and $\lambda=1$, we find that
$\frac{R_{circ}}{a}=0.20$.  This value is about the same as that for
standard Roche-lobe overflow, so the disk properties should be
essentially the same for such a system as for a Roche lobe overflower
of the same orbital period.  If the wind speed is larger than the
escape velocity from the the donor star, then the circularization
radius will decrease.  This effect is potentially quite substantial,
even for small changes in the wind speed, given the $\lambda^{-4}$
dependence of the circularization radius.  However, it is not clear
that a decrease in the circularization radius will have a major effect
on the observed properties of the outbursts from these objects, since
viscous spreading should cause the disk to spread outwards until it
reaches the tidal truncation radius regardless of the circularization
radius; as long as the disc circularizes before the gas reaches the
accretor, the properties of the disk should be largely unaffected by
the circularization radius.  Essentially, the systems should have
slightly lower peak luminosities and shorter outbursts than systems at
orbital periods just longer than systems just above the maximum period
for the period gap, and slightly longer, brighter outbursts than those
of systems with periods just shorter than the period at which systems
leave the period gap.

\subsection{Should the systems be transient?}

We can now verify that the observed systems should, in fact, be
transients.  The mean accretion rate needed to sustain a persistent
accretion disk with an outer radius of $2\times10^{10}$ cm is about
$2\times10^{15}$ g/sec (Dubus et al. 1999), a factor of about 300
higher than the $10^{-13} M_\odot$/yr which represents the maximum
value that could be expected from period gap systems.  Even the
highest accretion rate systems should be below the accretion rate
where disks are stably ionized.  We can also check the converse,
whether the systems might be stably neutral (see e.g. Menou et
al. 1999).  To avoid the stable ``cold branch'' solution, the disk
truncation radius must be less than $5\times10^8$ cm for a mass
transfer rate of $10^{12}$ g/sec.  The Alfven radius for a neutron
star with a magnetic field of $10^8$ G will be about $6\times10^{6}$
cm for this accretion rate, so the disk can be reasonably expected to
extend inward far enough to allow for a hot state to develop.

\subsection{Expected versus observed number of sources}

The next issue is the detected number of very faint X-ray transients.
While we can make a clear case that period gap binaries would make a
plausible argument for producing VFXTs, the argument is rather
uninteresting if the expected number of objects is so low that only a
tiny fraction of the VFXTs could be produced through this mechanism.
The bulk of the known VFXTs are found in monitoring observations of
the Galactic Centre region, where mass segregation may lead to heavy
stars settling into the central parsec (e.g. Muno et al. 2005; Freitag
et al. 2006; Alexander \& Hopman 2009), and dynamical formation of
X-ray binaries may take place (Voss \& Gilfanov 2007). In this region,
one might expect to see an excess of X-ray binaries in general, but
especially of the classes of X-ray binaries which are oldest (and
hence have had the longest amount of time to segregate inwards).
Perhaps most importantly, at least in the very centre of the Galaxy,
dynamical interactions may form substantial numbers of X-ray binaries;
furthermore, it has been suggested that the dynamically formed
binaries may preferentially be formed by tidal capture of low mass
($M$$\ltsim 0.3 M_\odot$) donor stars (Voss \& Gilfanov 2007).  The
dynamical effects, and other uncertainties in binary stellar evolution
will allow us only to make very crude estimates of the number of VFXTs
expected from this scenario.  At the same time, the observational
constraints on the number of such systems are also not yet very well
defined.  Still, we can ensure that all the assumptions needed to get
the numbers of predicted and observed sources to match are plausible.

Regardless, we can estimate the number of outbursting sources that
would be expected by looking at the duty cycles, recurrence timescales
and numbers of the sources.  The lifetime of cataclysmic variables in
the period gap is about 200 million years (Spruit \& Ritter 1983), and
a similar duration should be expected for neutron star LMXBs in the
period gap (Podsiadlowski et al. 2002), This duration is comparable to
that of the bright phase of LMXB evolution which can be estimated
simply by dividing the mass of a typical donor star by the mass
accretion rate needed to sustain a bright LMXB, assuming that most
neutron stars accrete their entire $\sim1 M_\odot$ companions, and
evolve through the period gap.  By taking this approach, we can
circumvent many of the uncertainties in binary population synthesis
modeling.  

Therefore, the number of active ($L_X \gtsim 3\times10^{37}$ erg/sec)
neutron star X-ray binaries in the Galaxy should be roughly equal to
the total number of neutron star X-ray binaries in the period gap,
since a donor star of 1 $M_\odot$ will be entirely accreted in 200
Myrs to produce that luminosity. If LMXBs with small initial masses
(e.g. of $\sim0.4 M_\odot$) are relatively common, or if
non-conservative mass transfer dominates for bright neutron star X-ray
binaries, or if the masses of the companions to neutron stars are
still a substantial fraction of the initial masses at the time
accretion stops, the total number of period gap systems could actually
dominate by number over the bright systems observed in any one epoch.
For the Milky Way, Grimm et al. (2002) estimate that there are
typically about 30 active low mass X-ray binaries with luminosities of
at least $10^{37}$ ergs/sec, so one can safely assume that the Galaxy
has at least 20 period gap X-ray binaries.  We emphasize again, that
this number could increase by a factor of a few if the typical amounts
of mass accreted over the lifetimes of bright X-ray binaries are
significantly less than 1 $M_\odot$.  Since 1 $M_\odot$ is the upper
bound for the mass of the donor stars for LMXBs with neutron star
accretors, and many of the LMXBs are even a bit brighter than
$3\times10^{37}$ ergs/sec, and hence have lifetimes shorter than 200
Myrs, this number is thus a conservative lower limit.

The ratio of the number of VFXTs to bright LMXBs could also rise if
mass transfer is non-conservative in the LMXB phase.  Strong disk
winds driven by irradiation and/or magnetic fields have been seen when
the accretion disks are bright and in soft states (e.g. Neilsen \& Lee
2009; Ponti et al. 2012).  In the black hole system A0620-00, a
mismatch between the brightness of the hot spot, which indicates the
mass transfer rate, and the mean integrated X-ray luminosity, has also
been noted as tentative evidence that a substantial fraction of the
mass is lost due to disk winds (Froning et al. 2011), but also
suggested that these winds could be driven in quiescence
(e.g. Blandford \& Begelman 1999).  It would not be unreasonable for
the ``effective lifetime'' of bright LMXBs to drop by a factor of a
few due to mass loss from disk winds in outburst, and for this mass
loss to have important implications for both the millisecond pulsar
birthrate problem (e.g. Kulkarni \& Narayan 1988) and for the fact
that the typical millisecond pulsar seems to have accreted only
$\sim0.2 M_\odot$ (e.g. Zhang et al. 2011), but clearly more work is
needed before this can be stated definitively.  All these concerns
taken together it seems reasonable that the number of period gap X-ray
binaries in the Galaxy at the present time is should not be less than
$\sim30$.  At the same time, the number can reasonably be expected to
exceed a few hundred only if the rate of LMXB formation was
considerably larger in the past than it is now.

If the typical outburst exponential decay timescale for the VFXTs is
about one week, and the peak luminosity is typically about $10^{35}$
ergs/sec, then with radiatively efficient accretion, one can estimate
that about $10^{-13} M_\odot$ is accreted per outburst.  For objects
at the high end of the range of wind mass transfer rates given above
(i.e. $\dot{M}\sim {\rm few} \times 10^{-14} M_\odot$/yr, the outburst
recurrence timescale should be every few years, consistent with what
is seen from the most frequently recurring VFXTs (e.g. Degenaar et
al. 2012).  

Next, we can use the variability survey of Degenaar et al. (2012) to
make a rough estimate of the number of VFXTs in the Galaxy.  This
survey covered the central 1.2 square degrees of the Galaxy with 10
epochs of data obtained over the course of 4 years.  Each epoch
reached a flux level corresponding to a luminosity of $5\times10^{33}$
erg/sec if at the Galactic Center.  If we assume that the typical VFXT
will have a peak luminosity of $10^{35}$ erg/sec, and an exponential
decay with a timescale of a few weeks, then these systems should
remain detectable for a few weeks, and most sources which undergo
outbursts should have been detected in the 10 epochs spread over 4
years.  Degenaar et al. (2012) reported the detection of 10 VFXTs in
their survey, so the total number in the central 1.2 square degrees of
the Galaxy is unlikely to be significantly more than about 20 --
Degenaar et al. (2012) reached similar conclusions on the basis of the
fact that no new transients were detected during this survey (most of
the region had already been surveys e.g. by Muno et al. 2006), and
many had been seen as recurring transients.

Next we must extrapolate the number of VFXTs seen in the central
region of the Galaxy to the Galaxy as a whole.  We can take the Bulge
model of Kent (1992), and integrate out the Bulge luminosity to a
projected radius of 40 pc, getting a total luminosity of about
$4\times10^8 L_\odot$, which is about 2\% of the total luminosity of
the Milky Way.  We can thus expect that there should be $\sim$500
VFXTs in the Galaxy.  This number is a factor of only a few larger
than the crude population estimate above.  Furthermore, the
observations suggest that there must really be some dynamical
enhancement of the VFXT numbers; four of the ten VFXTs in the sample
of Degenaar et al. (2012) are located within the central 25'' of the
Galaxy, which contains only about 1
region.  One of those objects is likely an edge-on system with a 7.9
hour period (see below for more discussion), but the other three may
very well be dynamically formed systems as in the scenario presented
by Voss \& Gilfanov (2007).  Ignoring the central 2 pc of the Galaxy
yields excellent agreement between predicted and observed numbers of
systems.  It does require that the accretion rates be near the upper
end of the possible range of accretion rates -- but the fact that many
of the systems are observed to be recurrent transients already
requires that; and it does require that the typical neutron star in a
LMXB accretes no more than about 0.2 $M_\odot$ -- but there are
indications for this already from the estimated masses of neutron
stars and the millisecond pulsar birth rates, and mechanisms for
explaining this in terms of disk winds, and both the alternative
evidence for low total amount of accretion and mechanisms for reducing
this amount from the total masses of the typical donor stars are
independent of the motivations of this paper.

\section{The effect of black hole accretors} 
The shortest period dynamically confirmed black hole system known, XTE
J1118+480, shows a peak X-ray luminosity of about $1.5\times10^{36}$
erg/sec, just above the luminosities of the very faint X-ray
transients, and has an orbital period of about 4.1 hours, just longer
than the period in the period gap.  Given that this source remained in
a low/hard state for its entire outburst, its low peak luminosity may
be partly due to its low radiative efficiency, if, e.g. the hard state
is well described by an advection dominated accretion flow
(e.g. Narayan \& Yi 1994) -- black hole systems with such short
orbital periods should all be expected to be below the 2\% of the
Eddington luminosity where systems are generally found in low hard
states (Maccarone 2003), and said radiative inefficiency should set
in.

One can then assume that $L_X \propto \dot{m}
\left(\frac{\dot{m}}{0.02\dot{m}_{EDD}}\right)$, where $\dot{m}_{EDD}$
is the mass accretion rate needed to get a source up to the Eddington
luminosity for radiatively efficient accretion.  The critical
accretion rate for triggering an outburst scales as $R$, or as
$P_{orb}^{2/3}$, assuming that the ratio of the circularization radius
to the orbital separation is a weak function of the orbital period
(e.g. Frank, King, Rane 2002).  The critcial X-ray luminosity will
then scale as $P_{orb}^{4/3}$.  We then can take XTE~J1118+480 as a
typical example, and find that black hole X-ray binaries with orbital
periods less than 3.6 hours will peak at X-ray luminosities below
$10^{36}$ erg/sec, making them very faint X-ray transients.  The duty
cycles of the outbursts of such systems will also be lower than
inferred from assuming that the systems are radiatively efficient by
factors of a few, meaning that the inferred mass transfer rates become
possible in the context of normal binary evolution.  On the other
hand, black hole X-ray binaries cannot explain VFXTs with Type I
bursts (e.g. Del Santo et al. 2007), so one cannot explain the whole
VFXT population as black hole X-ray binaries in radiatively
inefficient spectral states.

\section{Possible tests of the scenario}

The most obvious test of this mechanism obviously comes from measuring
the orbital periods of the systems and seeing that the are in the
period gap.  The observed period gap is from 2.1 to 3.1 hours for
cataclysmic variables (Knigge 2006).  Such measurements can be made
for in the optical or infrared from ellipsoidal modulations, but the
volume out to which they are feasible is rather small, given the
faintness of the donor stars.  It would not be surprising if a
substantial fraction of the period gap accretors turn out to be
accreting millisecond X-ray pulsars.  Indeed, of the two known X-ray
binaries that have been measured to fall into this range of periods,
one is IGR~J00291+5934 (Galloway et al. 2005), an accreting
millisecond pulsar.  Galloway et al. (2005) argue that that object is
likely to be fed from a sub-stellar mass donor in Roche lobe contact,
in part on the basis of the requirement of a very low inclination
angle needed to allow a main sequence hydrogen-rich donor star. The
other is a black hole system, MAXI~J1659-152, which has been suggested
to have a donor star made mostly of helium, which would allow the star
to maintain Roche lobe contact even at a short orbital period
(Kuulkers et al. 2012).

It should be noted that both of these systems have periods within the
empirical period gap found for cataclysmic variables.  They are at
shorter periods than those expected in the calculations of
Podisadlowski et al. (2002) for X-ray binaries which are unevolved at
the start of mass transfer, but in the period gap for some ranges of
systems which are partially evolved at the start of mass transfer.

One very faint X-ray transient has been reported to show suggestive
evidence of an orbital period of 7.9 hours (Muno et al. 2005).  This
system does not refute the scenario for producing very faint X-ray
transients we present here.  Firstly, Muno et al. (2005) note that the
integration time on this object is only 14 hours, less than 2 cycles
of the putative orbital period, and thus they urge caution in
interpreting the data.  Secondly, they note that if the periodicity is
real, it is likely to come from an eclipse.  Eclipsing X-ray binaries
are usually ``accretion disk corona'' sources, where the observed
X-ray luminosity is only 1-10\% of the intrinsic luminosity
(e.g. Bayless et al. 2010).  Transient ADC sources may represent a
fraction of the VFXTs, but are unlikely to represent the bulk of them
-- since geometric considerations indicate that there should be
several transients observed at lower inclinations angles for every
transient ADC source.

\subsection{Optical peak brightness of VFXTs}
An additional test is that the outbursts of the very faint X-ray
transients should be faint in the optical and near-infrared bands
(with the near-infrared likely being more important observationally
because searches for these objects are dominated by monitoring
campaigns on heavily extincted regions near the Galactic Center).
Taking the $V$-band luminosity to be $L_V$, one expects that $L_V
\propto L_X^{1/2} R$ if the optical flux is dominated by reprocessing
of X-rays which illuminate the outer accretion disk, and it has been
shown that this relation holds up well when compared with real data
(van Paradijs \& McClintock 1994; Russell et al. 2006).  This method
provides only crude orbital period estimates, especially when combined
with uncertainties in source distance and whether the accretor is a
black hole or a neutron star.  The one VFXT for which good
simultaneous optical and X-ray data have been obtained, XTE J1719-291,
has X-ray and optical luminosities consistent with those seen from the
2.0-hour period system SAX~J1808-365, provided the source is at a
distance of 8 kpc (Armas-Padilla et al. 2011).  This test is only a
weak one, since the differences between black holes and neutron stars
can lead to large differences in the inferred orbital period for a
system, and differences in inclination angle can lead to substantial
uncertainties, since at low luminosities, the X-rays are likely to
come from relatively large scale height regions, while the optical
emission necessarily comes from a thin disk.  At low luminosities, the
donor star's optical emission and jet emission can also become
important.  Nonetheless, any very faint X-ray transients which can be
confirmed (e.g. by bursting) to be neutron stars should have
relatively faint peak outburst fluxes in the optical and infrared.

\subsection{Superhumps}
In the context of this model, the very faint X-ray transients should
all be good candidates to show superhumps in outburst.  Superhumps are
periodic variations with timescales a few percent longer than the
orbital period of the binary system, and are seen in binaries with
mass ratios less than about 0.35 (Patterson et al. 2005).  The origin
of the periodicity is thought to be due to the growth of an
eccentricity in the outer accretion disk (e.g. Whitehurst 1988).
Since period gap systems with neutron star accretors will always have
mass ratios less than about 0.18, they are comfortably within the
range of mass ratios where superhumps should be seen.  It may thus be
that the regular dwarf nova type outbursts produce VFXT-like
outbursts, and the superoutbursts, which show superhump behaviour,
produce the brighter outbursts from systems with similar orbital
periods which have been seen already with all-sky instruments.

\section{Discussion and conclusions}

We have presented a new scenario for producing very faint X-ray
transients -- that they are X-ray binaries in the period gap, so that
the systems are (barely) detached binaries, in which accretion takes
place from the stellar winds of the M-dwarf donor stars.  Almost no
observational data currently exists on the orbital periods of the
VFXTs, but this is, in fact, in agreement with the expectations that
these systems will be optical faint both in quiescence and in
outburst.

\section{Acknowledgments}

We thank Christian Knigge, Rudy Wijnands, Nathalie Degenaar, Ed van
den Heuvel, Brian Warner, Thomas Tauris, Joey Neilsen and Yang Yi-Jung
for useful discussions.  We also thank an anonymous referee for a
brief, but constructive set of comments which have helped improve this
paper.

\label{lastpage}

\end{document}